\documentclass[11pt]{article}

\usepackage[T1]{fontenc}
\usepackage{times}
\usepackage[letterpaper,margin=1in]{geometry}
\usepackage{authblk}
\usepackage{natbib}
\usepackage{hyperref}
\usepackage{url}
\usepackage{booktabs}
\usepackage{amsmath}
\usepackage{graphicx}
\usepackage{xcolor}
\usepackage{enumitem}
\usepackage{longtable}
\usepackage{float}

\usepackage{amsmath,amsfonts,bm}

\def\eqref#1{equation~\ref{#1}}

\def\1{\bm{1}}

\def\vmu{{\bm{\mu}}}

\def\vc{{\bm{c}}}

\def\ve{{\bm{e}}}

\def\vh{{\bm{h}}}

\def\vu{{\bm{u}}}

\def\vx{{\bm{x}}}

\def\vz{{\bm{z}}}

\def\mH{{\bm{H}}}

\def\mP{{\bm{P}}}

\def\mR{{\bm{R}}}
\def\mS{{\bm{S}}}

\DeclareMathAlphabet{\mathsfit}{\encodingdefault}{\sfdefault}{m}{sl}
\SetMathAlphabet{\mathsfit}{bold}{\encodingdefault}{\sfdefault}{bx}{n}

\newcommand{\R}{\mathbb{R}}

\title{Learning Interpretable Tumor Microenvironment Representations \\ by Fitting
Pan-Cancer Cell State--Niche Correlation}

\author[1]{Xiao Xiao\thanks{Corresponding author.}}
\author[2]{Jiashu He}
\author[1]{Shiyang Zhang}
\author[3]{Marissa Mao}
\affil[1]{Yale University}
\affil[2]{University of Pennsylvania}
\affil[3]{University of Michigan, Ann Arbor}

\date{}

\begin{document}

\maketitle

\begin{abstract}
In the tumor microenvironment, a cell's state is influenced by cell--cell interactions (CCIs) with neighboring cells in its niches. Identifying dysregulated CCIs that are associated with pathogenic processes pinpoints targets for drug discovery. Imaging-based spatial transcriptomics and single-cell RNA sequencing provide, respectively, single-cell spatial information and transcriptome-wide measurements needed to study CCIs, but neither modality provides both. Existing spatial transcriptomics foundation models also cannot effectively learn from spatially resolved single-cell data with full-transcriptome coverage, explicitly infer the CCI mechanisms driving cell state--niche associations, or be interpretable enough to support direct biological interpretations. Here, we present GITIII-scale, a hierarchical, interpretable pan-cancer spatial transcriptomics foundation model for TME representation learning that investigates cell state--niche associations and their underlying ligand--receptor (LR) signaling pathways. GITIII-scale uses transformers to model interactions between pairs of cells at defined spatial distances, an interpretable single-layer graph transformer without a feed-forward network to decompose how each gene in a receiver cell is influenced by each neighboring sender cell, and a graph transformer to generate cellular-neighborhood embeddings. Trained on our assembled pan-cancer database of specimen-matched scRNA-seq and imaging-based spatial transcriptomics datasets, GITIII-scale generated TME embeddings that recovered niche-associated state changes more accurately than existing spatial transcriptomics foundation models in cancer types unseen during training. A case study of an unseen breast cancer dataset further demonstrated the model's interpretability by identifying potentially drug-targetable LR pathways associated with endothelial overgrowth and tumorigenesis.
\end{abstract}

\section{Introduction}

Cells coordinate their behavior through cell--cell interactions within their local
microenvironment. A CCI depends jointly on ligand--receptor (LR) signaling, the physical
distance between the two cells, and the downstream regulatory network of the receiver, producing a
context-dependent mapping from a cell's niche (neighboring cells) to its state. We refer to this mapping as the \emph{cell
state--niche correlation}: how much of a cell's deviation from the mean expression level of its cell
type is attributable to the cells around it. In tumors, this quantity is crucial for disease:
different arrangements of malignant, immune, stromal, and endothelial cells induce distinct
pathological programs \citep{hanahan2022hallmarks,devisser2023tme}. For instance, tumor-derived vascular endothelial growth factor (VEGF) acts on receptors
on neighboring endothelial cells to drive their proliferation and sprouting, and the
resulting angiogenesis supplies the growing tumor \citep{ferrara2003vegf}. 

Quantifying cell state--niche correlations and identifying the specific LR pathways that drive them are therefore crucial for understanding cancer dynamics and developing targeted therapies. For example, the endothelial overgrowth state was found to correlate with malignant cells in
the niche through VEGF signaling, which led to the development of VEGF-targeted drugs such as bevacizumab
\citep{ferrara2004bevacizumab} that suppress tumor growth by inhibiting tumorigenesis induced by VEGF signaling. Learning cell
state--niche correlation across cancer types can further distinguish conserved microenvironmental
programs from cancer-specific interactions, prioritize robust therapeutic targets, support
drug repurposing, and reveal the cellular contexts in which a target is most likely to be
effective.

Comprehensively modeling the tumor microenvironment (TME) requires spatial context and
transcriptome-wide measurement at once, and no single assay provides both. Single-cell RNA
sequencing (scRNA-seq) reads full transcriptomes but dissociates the tissue and discards
position. Imaging-based spatial transcriptomics (IST) preserves single-cell position but
measures a targeted panel of 312 to 6{,}175 genes across the platforms used here
\citep{chen2015merfish,janesick2023xenium}. Neither technique alone captures both the niche
and the state it induces, but profiling the same specimen with both does, which is the
resource we build here, with the two modalities integrated by SpaIM \citep{li2025spaim}.

Spatial transcriptomics foundation models
\citep{blampey2025novae,schaar2025nicheformer,madhu2025heist,wang2025scgptspatial,wang2026spatialformer,birk2026terra} have
recently demonstrated that large-scale pretraining captures transferable patterns of tissue
organization and cellular context. Their objectives, however, target tissue architecture,
spatial domains, cell identity, or expression similarity, none of which reveals cell
state--niche correlation: different cell types may occur within the same domain, and cells with
different niches and different states may also occur within the same domain. A model can
therefore be perfectly domain-discriminative while carrying no information about how the
niche shifted the receiver's state, and its TME representation cannot be decomposed into
directional CCIs or the pathways that would make it useful for target discovery. Existing foundation
models are also trained on data that carries partial biological measurement: either on spot-level data, where several cells
per spot blur the correlation, or on IST data without transcriptome-wide measurement, so
only a panel-sized slice of the biology is visible
\citep{hu2021spagcn,long2023graphst,madhu2025heist,blampey2025novae}, a dilemma that
paired IST and scRNA-seq on one specimen removes. Finally, previous deep learning-based CCI analysis methods such as \citep{xiao2025gitiii,li2023holonet,fischer2023ncem} are designed to be dataset-specific models and have not been scaled up or trained on a large corpus to learn generalizable CCI features; Appendix~\ref{sup:vsgitiii} details how GITIII-scale differs from its closest predecessor.

\begin{figure}[t]
\centering
\includegraphics[width=\textwidth]{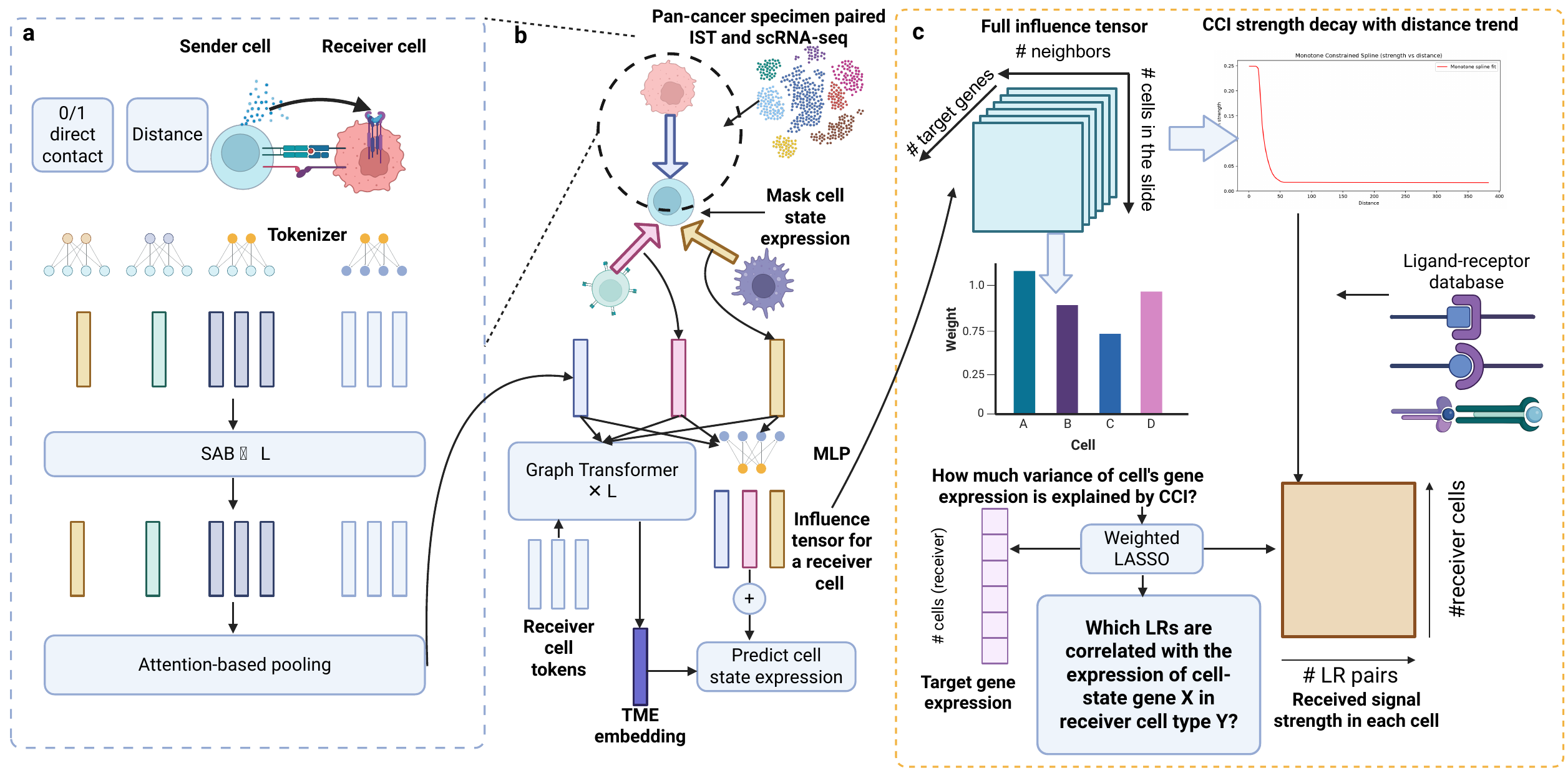}
\caption{GITIII-scale model architecture. \textbf{a.} GITIII-scale treats the gene expression of receiver and sender cells, the distance, and direct contact features between one sender and one receiver cell as tokens, and uses transformer blocks to model them. \textbf{b.} GITIII-scale utilizes a single-layer graph transformer block without a feed-forward network to retain interpretability and estimate additive CCI influence from different sender cells to the receiver cell, then uses a multi-layer transformer block to generate TME embedding and estimate higher-order CCI influence from different sender cells to the receiver cell. \textbf{c.} The interpretability analysis pipeline of GITIII-scale. A LASSO regression is used to identify the LR pair(s) whose signaling strength is correlated with target gene expression in receiver cells. LR signaling strength is estimated by modeling the decay of interaction strength with distance and integrating information from the LR database, with receiver cells weighted by their cell-state prediction accuracy in the regression.}
\label{fig:overview}
\end{figure}

To this end, we introduce GITIII-scale (graph inductive bias transformer for intercellular interaction investigation, scaled), a hierarchical interpretable pan-cancer foundation model for TME representation learning to investigate cell state--niche correlation and its underlying LR pathways (Figure~\ref{fig:overview}). At the first level, sender and receiver
cells and intercellular distances are tokenized; the cross-attention module then models directional
interaction, and an interpretable single-layer graph inductive-bias
transformer takes the interaction tokens and estimates the influence of each sender cell on each gene of the receiver. These
influences are additive, so the summed contributions of a receiver's senders constitute its
predicted state and every prediction decomposes exactly by sender and by gene. At the
second level, a multi-layer graph transformer takes the resulting sender-receiver pair interaction representations to
model higher-order cell state--niche correlation and produces an embedding of the niche as
a whole. GITIII-scale is self-supervised pretrained on biological sample-matched scRNA-seq and IST data
comprising $4{,}254{,}069$ spatial cells, imputed to a common 20{,}000-gene vocabulary, and is evaluated on held-out cancer types and platforms to test whether its
TME embeddings recover cell state--niche correlation under biological distribution shift. Overall,
we propose GITIII-scale and make the following contributions:

\begin{itemize}
\item \textbf{Foundation model for CCI representation learning in TME trained on a new pan-cancer corpus.} GITIII-scale is the first interpretable foundation model to learn how cell states are influenced by their niche, enabling CCI modeling and representation learning. We assemble and release a pan-cancer corpus of sample-matched imaging-based spatial transcriptomics and scRNA-seq from the same biological specimens, enabling the model to learn richer information from imputed full transcriptomics at single-cell resolution compared to previous models that either lose single-cell resolution or full transcriptomics measurement.

\item \textbf{Pan-cancer pretraining enables state-of-the-art performance in generating embeddings that represent cell state--niche correlations.} Pretrained to model how cellular niches shape receiver-cell states across cancer types, GITIII-scale can generate TME embeddings that recover niche-associated state changes more accurately than existing spatial transcriptomics foundation models on cancer types unseen during training. GITIII-scale also outperforms smaller models trained specifically on the target datasets.

\item \textbf{Interpretable foundation model for biological findings.}
Our model's interpretable architecture decomposes each predicted receiver-cell state into gene-level contributions from individual neighboring cells. This goes beyond conventional CCI analysis, which typically identifies candidate ligand--receptor events but does not model their downstream transcriptional consequences at the single-cell level. GITIII-scale also enables statistical testing of which ligand--receptor axes and which cell types are associated with specific gene programs in a receiver cell type, supporting potential biological discovery and therapeutic-target prioritization.

\end{itemize}

\section{Method}
\label{sec:method}

\subsection{Data preprocessing}
\label{sec:setup}

\paragraph{Cell type / cell state split and graph construction.} Let $\vmu_c$ denote the mean expression profile of
cell type $c$ within a sample. We decompose each measured profile $\vx_i$ into a cell type component $\vmu_{t(i)}$, representing the average expression for cells of the same cell type, and
a cell state component $\vz_i$, representing the cell's deviation from its cell type average level,
\begin{equation}
\vx_i = \vmu_{t(i)} + \vz_i,
\qquad
\vz_i = \vx_i - \vmu_{t(i)},
\label{eq:split}
\end{equation}
We then construct a $k$-nearest-neighbor graph and a Delaunay
triangulation over the cell centroids $\mP$. The Delaunay triangulation indicates the neighboring cells that directly contact the central cell. The \textbf{cellular neighborhood}
$\mathcal{N}_i$ of a receiver cell $i$ comprises itself and its $k = 49$ nearest cells together with
their types, states, and relative geometry, and the Delaunay graph identifies which of them
are in direct physical contact with $i$.

\paragraph{Same-type masking.} Neighboring cells of the same type tend to be in similar states simply because they share similar niche features, and a model given access to neighboring cells of the same cell type as the central cell to predict could reach high accuracy
by copying them, thus learning nothing about cell--cell communication \citep{xiao2025gitiii}. As a result, for cells within $\mathcal{N}_i$ that share the receiver's type,
including the receiver itself, their expressions are masked as $\vmu_t$ with their state removed in training.

\subsection{Model architecture}
\label{sec:arch}

\paragraph{Tokenizing cells and distances between cells.} Treating
genes as tokens, as language-inspired single-cell models do, yields sequences of tens of
thousands of elements per cell and makes a neighborhood of fifty cells computationally
prohibitive; it also places the attention mechanism between genes, when the relationships of
interest here are between cells. GITIII-scale therefore treats \emph{cells} as tokens: two multi-layer perceptrons (MLPs), one for sender cells and one for the receiver cell, map a
$G$-dimensional gene expression profile to $T = 32$ tokens of width $d = 256$, so that a cell is represented differently depending on whether it is sending or
receiving:
\begin{equation}
\mS_j = \mathrm{reshape}\big(\mathrm{MLP}_{\mathrm{snd}}(\vx_j)\big),
\quad
\mR_i = \mathrm{reshape}\big(\mathrm{MLP}_{\mathrm{rec}}(\vx_i)\big)
\quad \in \R^{T \times d},
\qquad T = 32,\; d = 256 .
\end{equation}
The two roles are kept separate because the biology is asymmetric: the genes that matter for
sending signal are ligand genes, while those that matter for receiving
it are receptors and downstream regulatory genes, and a shared encoder would not project them to different spaces.

The pairwise spatial distance $d_{ij}$ is expanded into a small set of decay
features spanning contact-range to diffusive length scales (Appendix~\ref{sup:distance}) and mapped
by an MLP to a single distance token $\vu_{ij} \in \R^{d}$, while a contact token
$\vc_{ij} \in \R^{d}$ embeds whether two cells are in direct contact according to the Delaunay triangulation.

The sender cell tokens, distance token, and the direct contact token of one sender-receiver pair are then processed through self-attention blocks (SAB), and the resulting token set is concatenated with the receiver cell tokens and processed by another stack of SABs (Figure~\ref{fig:overview}a). The outputs of the SABs are then pooled by a multi-head cross-attention block (MCAB)
\citep{jaegle2021perceiver}, in which $L = 4$ learned latent queries attend to the tokens
and return a fixed-size, permutation-invariant summary that is flattened and projected:
\begin{align}
\tilde{\mS}_{ij} &= \mathrm{SAB}\big([\,\mS_j;\, \vu_{ij};\, \vc_{ij}\,]\big)
  \;\in\; \R^{(T+2) \times d}, \\
\mH_{ij} &= \mathrm{SAB}\big([\,\tilde{\mS}_{ij};\, \mR_i;\, \vu_{ij};\, \vc_{ij}\,]\big)
  \;\in\; \R^{(2T+4) \times d}, \\
\ve_{ij} &= W_{c}\,\mathrm{flatten}\big(\mathrm{MCAB}(\mH_{ij})\big),
  \qquad \mathrm{MCAB}(\cdot) \in \R^{L \times d},\;\; \ve_{ij} \in \R^{256},
\label{eq:pair}
\end{align}
with three SAB blocks at each stage.

\paragraph{Level one: an interpretable additive graph inductive-bias transformer.} Both prediction heads take the sender, receiver, distance, and direct contact tokens produced above (Figure~\ref{fig:overview}b) as inputs. For the first head,  similar to the interpretability analysis in GITIII \citep{xiao2025gitiii}, two MLPs each map
$\ve_{ij}$ to a per-gene attention logit and a per-gene value, and attention scores are normalized
\emph{across senders, independently for each gene}. Then we multiply the value by attention scores and sum them up to directly predict the cell state expression:
\begin{equation}
\alpha_{ij}^{(g)} =
\frac{\exp\big(W_A(\ve_{ij})^{(g)}\big)}{\sum_{j'}\exp\big(W_A(\ve_{ij'})^{(g)}\big)},
\qquad
I_{ij}^{(g)} = \alpha_{ij}^{(g)}\, W_V(\ve_{ij})^{(g)},
\qquad
\hat{z}_i^{(1)(g)} = \sum_{j=1}^{k} I_{ij}^{(g)} .
\label{eq:influence}
\end{equation}

Equation~\ref{eq:influence} is where the interpretability of the model is established:
(1) each entry of the influence tensor $\mathcal{I} \in \R^{N \times k \times G}$, is computed from
the features of cells $i$ and $j$ and their distance using a one-layer aggregation, so no multi-layer graph neural network (GNN) can
distort it; (2) $\sum_{j=1}^{k} I_{ij}^{(g)}$ directly predicts the cell-state expression target, which is not distorted by a non-linear transform that would destroy interpretability. As a result, each entry $I_{ij}^{(g)}$ states how much
of the receiver cell $i$'s gene $g$'s predicted cell state (gene expression deviation from its cell type average expression) is attributable to its neighbor $j$. The mathematical basis for this interpretation, including why the influence tensor provides an exact additive decomposition of the model prediction, is detailed in Appendix~\ref{sup:interpretableproof}.

\paragraph{Level two: higher-order effects and the TME embedding.} Additivity makes the first head interpretable, but also limits it. A sum of influence from all sender cells of a receiver cell
cannot express the effect that requires a combination of neighbors, such as a signal that
acts only when a second cell type is co-present, or a suppression that happens only in the
absence of another type of cell. Rather than sacrifice the additive decomposition, we add a second head
that reads the same tokens as the level-one head and makes them attend to each other before pooling:
\begin{equation}
\vh_i = \mathrm{flatten}\Big(\mathrm{MCAB}\big(\mathrm{SAB}
  ([\,W_p\ve_{i1},\dots,W_p\ve_{ik};\, \mR_i\,])\big)\Big) \in \R^{L \cdot d} = \R^{1024},
\qquad
\hat{\vz}^{(2)}_i = W_{\mathrm{out}}\,\vh_i,
\end{equation}
with four SAB blocks. The pooled representation $\vh_i$ is the TME embedding. 

\subsection{Self-supervised objective functions} 
We want to model whether a gene's variation across cells in the same cell type is correlated with variation in their
neighborhoods. The absolute scale of the cell state expression is largely governed by panel sensitivity and batch effects, which differ across platforms and would make the mean squared error loss reflect assay properties rather than biology. We
therefore optimize per-gene Pearson correlation computed \emph{across cells within a gene in the same cell type},
which directly measures cell state--niche
correlation. Over a set $\mathcal{B}$ of receiver cells of the same receiver cell type in the same sample,
\begin{equation}
\rho^{(g)} = \mathrm{corr}_{i \in \mathcal{B}}\big(\hat{z}_i^{(g)},\, z_i^{(g)}\big),
\qquad
\mathcal{L}_{\mathcal{B}} = \sum_{g} w^{(g)} \Big[\big(1 - \rho^{(g)}\big)
   + \lambda \big\|\tilde{\hat{z}}^{(g)} - \tilde{z}^{(g)}\big\|^2 \Big],
\label{eq:loss}
\end{equation}
where the weights $w^{(g)} \propto \mathrm{std}_{i \in \mathcal{B}}(z_i^{(g)})$, normalized
to mean one to make genes with more intra-cell-type variance contribute more in the training. The second term is a mean squared error
between $z$-scored predictions and targets with $\lambda = 0.05$ to stabilize the training. The training loss sums Equation~\ref{eq:loss} over the two heads. GITIII-scale has
$2{,}296{,}747{,}232$ ($\sim$2.3B) parameters; the full architecture and optimization
specification is given in Appendix~\ref{sup:hparams}.

\subsection{Pretraining corpus curation}
\label{sec:corpus}

To obtain the imputed full transcriptomics spatial transcriptomics data, we searched for biological specimen-paired IST and scRNA-seq datasets. This yields a pretraining corpus of $4{,}254{,}069$ spatial cells over $10$ samples spanning $4$ tumor contexts and $2$ imaging platforms. Appendix~\ref{sup:datasets} and Table~\ref{tab:accessions} summarize the samples we found and used in training and evaluating GITIII-scale, together with their
per-sample accessions and download links. Cell type annotations are taken from each dataset's original
study release. Every IST sample is paired with
scRNA-seq from the same specimen, which is what makes it possible to impute the measured
panel onto the shared $20{,}000$-gene vocabulary. The scRNA-seq and IST
integration and gene imputation are performed with the state-of-the-art SpaIM \citep{li2025spaim} model. The breast sample used in the case study in Section~\ref{sec:casestudy} is held out from pretraining.

\subsection{Interpretability analysis of inferring LRs that are associated with target genes in the receiver cell type}
\label{sec:attribution}

The influence tensor identifies the cell state of the receiver cell that is related to its niche features, but it does not
point out the signaling axis through which the effect was mediated, and related data-driven methods are lacking. Bridging that gap requires
external knowledge of ligand--receptor biochemistry, and the following procedure utilizes the
curated LR database \citep{jin2021cellchat,zhao2023neuronchat} to infer which LR is correlated with the influence. Given a receiver cell type $c$ and a
target gene $g^\ast$, we recover the responsible axes in the following steps
(Figure~\ref{fig:overview}c).

\textbf{(1) Distance scaler.} A monotone spline between the estimated influence from a sender cell to a receiver cell targeting one gene and the distance between the sender and the receiver cell is fitted to form the distance scaler $s_{c,g^\ast}(d)$. It is estimated
\emph{separately for each (receiver type, target gene) combination}, because the signal strength decay trend is different for the correlated LR targeting each (receiver type, target gene) combination: the LR that influences target gene A in receiver type X may be a contact-dependent program, while the LR that influences target gene B in receiver type Y may be
a cytokine-driven program with another distance decay trend. Thus
$s_{c,g^\ast}$ is the model's own learned interaction range for that specific readout, not a
global assumption that fits all of them.

\textbf{(2) Cell weights.} A cell whose cell state expression cannot be explained by the model is likely to carry less
information about how CCI influences cell state and may contain more noise, so it should be weighted less in the analysis. We therefore
weight each receiver by the fidelity of its own prediction,
\begin{equation}
\omega_i = \exp\!\big(-\,|z_i^{(g^\ast)} - \hat z_i^{(g^\ast)}| \,/\, \max_{i'} |z^{(g^\ast)} - \hat z^{(g^\ast)}|\big),
\end{equation}
so that cells whose expression is well explained by their neighbors contribute most.

\textbf{(3) LR signal strength.} For each LR pair $(\ell,r)$, we scale the ligand expression
of every sender by the distance scaler calculated from the distance between the sender and the receiver cell, then sum them over the receiver cell's neighboring sender cells, and multiply by
the receiver's own receptor expression to form the received signaling strength by the receiver cell:
\begin{equation}
S_i^{(\ell,r)} = \Big(\textstyle\sum_{j \in \mathcal{N}_i} s_{c,g^\ast}(d_{ij})\,
  \mathrm{gm}\big(\vx_j^{(\ell)}\big)\Big) \cdot \mathrm{gm}\big(\vx_i^{(r)}\big),
\end{equation}
where $\mathrm{gm}(\cdot)$ denotes the geometric mean over the subunits of a multimeric
ligand or receptor, so that a complex scores highly only when all of its components are
present. The product form encodes the requirement that signaling needs both an available
ligand nearby and a receptor on the receiving cell. The LR pairs are extracted from CellChat \citep{jin2021cellchat} and NeuronChat \citep{zhao2023neuronchat}.

\textbf{(4) Sparse selection.} The measured target $z^{(g^\ast)}$ is regressed on all
candidate $S^{(\ell,r)}$ by weighted LASSO, with the penalty $\alpha$ selected by
cross-validated mean squared error and cells weighted by $\omega_i$. Sparsity is essential
here: several thousand LR pairs are tested against one gene, many of them collinear through
shared subunits, and an unregularized fit would distribute the signal across all of them.
Axes with nonzero coefficients are retained and ranked by coefficient magnitude to select the most correlated LR.

\textbf{(5) Stability.} The selection is repeated five times, and only axes that are
top-ranked across all repeats are reported.

\section{Experiments}
\label{sec:experiments}

\subsection{OOD Evaluation datasets}
\label{sec:exp-data}

We evaluate on three held-out samples with specimen-matched IST and scRNA-seq data spanning two platforms and three cancer types:
the breast cancer Xenium FFPE sample \citep{janesick2023xenium}, and colorectal carcinoma and melanoma MERFISH
samples \citep{zhang2026spatialecotypes}. All three are withheld from
GITIII-scale pretraining in their entirety. To further establish the out-of-distribution (OOD) generalization of GITIII-scale, neither breast cancer nor the melanoma cancer type appears in the
training corpus, and the MERFISH platform is also absent from the training data. The comparison on the breast cancer dataset is
conservative for us and favorable to the baselines: these samples belong to the pretraining
corpora of both Novae \citep{blampey2025novae} and HEIST \citep{madhu2025heist}, so their
embeddings are in-distribution here while ours are not.

For each cell, we build a cellular neighborhood as described in Section~\ref{sec:setup}, and
every neighboring sender cell sharing the receiver's cell type, the receiver included, has its expression
replaced by that type's mean $\vmu_{t(i)}$. The mask is intended to make a fair comparison between methods and to truly evaluate whether different models can encode cell state--niche correlations. As discussed above, cell state is spatially autocorrelated within a type, so a probe with access to
the unmasked states of nearby same-type cells can score well by interpolating from them. As a result, the comparison does not
reward the embedding that preserves a cell's own expression most faithfully. The same
masked neighborhoods are supplied to every model with different data preprocessing procedures in the comparison.

\subsection{Evaluation Metrics}
\label{sec:exp-metrics}

Spatial foundation models differ in architecture, in pretraining corpus, and in the
downstream applications their authors provided. Previous foundation models do not enable zero-shot prediction of cell state--niche correlations either, which makes end-to-end comparison ambiguous. To enable comparison, we
compare all foundation models as \emph{frozen encoders} under a single downstream probe and train all dataset-specific models on each dataset individually, so
that differences in performance reflect what a representation encodes about cell
state--niche correlation. Each
model produces a TME embedding $\vh_i$ per cell from the same cell type expression-masked neighborhood, with per-model
extraction details given in Appendix~\ref{sup:extraction}. A two-layer MLP then predicts the cell's
state from that embedding together with its cell-type identity,
\begin{equation}
\hat{\vz}_i = \mathrm{MLP}\big([\,\vh_i;\, \mathrm{Emb}(t(i))\,]\big),
\end{equation}
using a $128$-dimensional cell-type embedding and a $1024$-unit hidden layer. We also repeated the experiments using a similar MLP with a $256$-unit hidden layer to remove the influence of different embedding size. Supplying the
cell type explicitly means the probe need not recover it from the embedding, so the
measurement isolates the niche information a representation adds beyond identity. Cells are
randomly split into training, validation, and test sets with ratio $0.7 / 0.15 / 0.15$; the probe is
trained on the training split, early-stopped on the validation split, and all reported
numbers are computed on the test split over five seeds. Architecture, split, and optimizer
are held identical across models, so that the model embedding is the only variable.

The two metrics focus on how much each model's output TME representation encodes the cell
state--niche correlation. The first metric is the cell-state Pearson correlation coefficient (PCC):
\begin{equation}
\mathrm{PCC} = \frac{1}{|\mathcal{G}|}\sum_{g \in \mathcal{G}}
  \mathrm{corr}_{i \in \mathrm{test}}\big(\hat z_i^{(g)},\, z_i^{(g)}\big),
\end{equation}
The
second is variance explained, which is also used by NCEM \citep{fischer2023ncem},
\begin{equation}
R^2 = \frac{1}{|\mathcal{G}|}\sum_{g \in \mathcal{G}}
  \left(1 - \frac{\sum_{i \in \mathrm{test}} \big(z_i^{(g)} - \hat z_i^{(g)}\big)^2}
  {\sum_{i \in \mathrm{test}} \big(z_i^{(g)} - \bar z^{(g)}\big)^2}\right),
  \qquad \bar z^{(g)} = \frac{1}{|\mathrm{test}|}\sum_{i \in \mathrm{test}} z_i^{(g)},
\end{equation}
In the benchmarking, we calculated the above-mentioned metrics in observed gene's expression (Measured) and all genes' (observed and imputed) expression (All). Both metrics are computed per gene across test cells and then averaged over the
genes with nonzero variance.

\subsection{GITIII-scale outperforms previous spatial transcriptomics foundation model and smaller dataset-specific models}
\label{sec:results}

We compare against three groups of baselines: pretrained spatial foundation models that produce
niche embeddings, namely TERRA \citep{birk2026terra}, HEIST \citep{madhu2025heist}, Novae
\citep{blampey2025novae}, and SpatialFormer \citep{wang2026spatialformer}; non-pretrained graph
models trained directly on each target dataset; and a probe built from neighborhood cell-type
composition alone (Appendix~\ref{sup:related}). For pretrained spatial foundation models, the input is the same cellular neighborhood with same cell type masking followed by each method's specific data preprocessing pipelines.

As shown in Table~\ref{tab:bench}, GITIII-scale achieves the best variance explained and the best mean PCC on all three held-out
samples. Which baseline is the strongest varies by sample (TERRA
\citep{birk2026terra} on breast and melanoma, HEIST \citep{madhu2025heist} on colorectal),
whereas the ordering of GITIII-scale is stable across cancer types and across both platforms.
NCEM with the graph convolutional network (NCEM-GCN) \citep{fischer2023ncem} and the graph attention network (GAT) \citep{velickovic2018gat} are not foundation
models and are trained with random initialization directly on each dataset's train/val/test split. The $k$-NN cell-type proportion probe, $\mathrm{MLP}$(cell-type embedding, niche
cell-type proportion), measures how much of the signal is available from neighborhood
composition alone; it recovers between $28\%$ and $54\%$ of GITIII-scale's variance explained
depending on sample and $k$. To test the influence of different MLP probes, we repeat the above-mentioned 10 methods $\times$ 3 datasets $\times$ 5 replicates $=$ 150 experiments on an MLP with a smaller hidden layer, whereas the results show that GITIII-scale still performs the best (Table~\ref{tab:bench256}).

\begin{table}[t]
\centering
\caption{Cell-state prediction on the three held-out samples ($1024$-unit probe). $R^2$ is variance explained, PCC the gene-averaged Pearson correlation, both on the test split; mean $\pm$ s.d.\ over five seeds. \emph{Measured} restricts targets to each platform's panel, \emph{All} uses every nonzero-variance gene. Best per column in bold.}
\label{tab:bench}
\scriptsize
\setlength{\tabcolsep}{3pt}
\begin{tabular}{l cc cc cc}
\toprule
 & \multicolumn{2}{c}{Xenium Breast} & \multicolumn{2}{c}{MERFISH CRC} & \multicolumn{2}{c}{MERFISH Melanoma} \\
\cmidrule(lr){2-3}\cmidrule(lr){4-5}\cmidrule(lr){6-7}
Method & $R^2$ & PCC & $R^2$ & PCC & $R^2$ & PCC \\
\midrule
\multicolumn{7}{l}{\emph{Measured genes}} \\
\textbf{GITIII-scale (ours)} & \textbf{0.168}\,{\scriptsize$\pm$0.003} & \textbf{0.398}\,{\scriptsize$\pm$0.004} & \textbf{0.059}\,{\scriptsize$\pm$0.000} & \textbf{0.243}\,{\scriptsize$\pm$0.003} & \textbf{0.112}\,{\scriptsize$\pm$0.002} & \textbf{0.328}\,{\scriptsize$\pm$0.004} \\
TERRA & 0.146\,{\scriptsize$\pm$0.001} & 0.370\,{\scriptsize$\pm$0.002} & 0.035\,{\scriptsize$\pm$0.002} & 0.198\,{\scriptsize$\pm$0.004} & 0.070\,{\scriptsize$\pm$0.002} & 0.268\,{\scriptsize$\pm$0.005} \\
HEIST & 0.124\,{\scriptsize$\pm$0.002} & 0.344\,{\scriptsize$\pm$0.003} & 0.041\,{\scriptsize$\pm$0.004} & 0.203\,{\scriptsize$\pm$0.005} & 0.072\,{\scriptsize$\pm$0.003} & 0.270\,{\scriptsize$\pm$0.009} \\
Novae & 0.111\,{\scriptsize$\pm$0.001} & 0.324\,{\scriptsize$\pm$0.001} & 0.031\,{\scriptsize$\pm$0.001} & 0.173\,{\scriptsize$\pm$0.006} & 0.047\,{\scriptsize$\pm$0.002} & 0.214\,{\scriptsize$\pm$0.006} \\
SpatialFormer & 0.062\,{\scriptsize$\pm$0.002} & 0.233\,{\scriptsize$\pm$0.005} & 0.012\,{\scriptsize$\pm$0.007} & 0.120\,{\scriptsize$\pm$0.006} & 0.009\,{\scriptsize$\pm$0.007} & 0.109\,{\scriptsize$\pm$0.018} \\
NCEM-GCN & 0.126\,{\scriptsize$\pm$0.003} & 0.351\,{\scriptsize$\pm$0.006} & 0.024\,{\scriptsize$\pm$0.001} & 0.159\,{\scriptsize$\pm$0.001} & 0.056\,{\scriptsize$\pm$0.003} & 0.240\,{\scriptsize$\pm$0.012} \\
GAT & 0.111\,{\scriptsize$\pm$0.004} & 0.324\,{\scriptsize$\pm$0.002} & 0.016\,{\scriptsize$\pm$0.000} & 0.132\,{\scriptsize$\pm$0.004} & 0.026\,{\scriptsize$\pm$0.012} & 0.173\,{\scriptsize$\pm$0.017} \\
$k$-NN prop.\ ($k{=}49$) & 0.081\,{\scriptsize$\pm$0.000} & 0.271\,{\scriptsize$\pm$0.001} & 0.026\,{\scriptsize$\pm$0.002} & 0.151\,{\scriptsize$\pm$0.004} & 0.059\,{\scriptsize$\pm$0.000} & 0.232\,{\scriptsize$\pm$0.003} \\
$k$-NN prop.\ ($k{=}25$) & 0.082\,{\scriptsize$\pm$0.001} & 0.272\,{\scriptsize$\pm$0.001} & 0.023\,{\scriptsize$\pm$0.001} & 0.140\,{\scriptsize$\pm$0.002} & 0.056\,{\scriptsize$\pm$0.001} & 0.224\,{\scriptsize$\pm$0.001} \\
$k$-NN prop.\ ($k{=}10$) & 0.084\,{\scriptsize$\pm$0.001} & 0.275\,{\scriptsize$\pm$0.001} & 0.018\,{\scriptsize$\pm$0.000} & 0.127\,{\scriptsize$\pm$0.004} & 0.044\,{\scriptsize$\pm$0.000} & 0.202\,{\scriptsize$\pm$0.001} \\
\midrule
\multicolumn{7}{l}{\emph{All genes}} \\
\textbf{GITIII-scale (ours)} & \textbf{0.193}\,{\scriptsize$\pm$0.003} & \textbf{0.433}\,{\scriptsize$\pm$0.003} & \textbf{0.067}\,{\scriptsize$\pm$0.000} & \textbf{0.261}\,{\scriptsize$\pm$0.001} & \textbf{0.120}\,{\scriptsize$\pm$0.002} & \textbf{0.340}\,{\scriptsize$\pm$0.004} \\
TERRA & 0.154\,{\scriptsize$\pm$0.001} & 0.387\,{\scriptsize$\pm$0.001} & 0.036\,{\scriptsize$\pm$0.002} & 0.201\,{\scriptsize$\pm$0.007} & 0.074\,{\scriptsize$\pm$0.002} & 0.274\,{\scriptsize$\pm$0.010} \\
HEIST & 0.127\,{\scriptsize$\pm$0.006} & 0.353\,{\scriptsize$\pm$0.009} & 0.041\,{\scriptsize$\pm$0.002} & 0.208\,{\scriptsize$\pm$0.005} & 0.071\,{\scriptsize$\pm$0.002} & 0.265\,{\scriptsize$\pm$0.001} \\
Novae & 0.119\,{\scriptsize$\pm$0.000} & 0.341\,{\scriptsize$\pm$0.000} & 0.033\,{\scriptsize$\pm$0.002} & 0.183\,{\scriptsize$\pm$0.004} & 0.049\,{\scriptsize$\pm$0.001} & 0.220\,{\scriptsize$\pm$0.002} \\
SpatialFormer & 0.051\,{\scriptsize$\pm$0.000} & 0.214\,{\scriptsize$\pm$0.000} & 0.017\,{\scriptsize$\pm$0.003} & 0.080\,{\scriptsize$\pm$0.017} & 0.010\,{\scriptsize$\pm$0.002} & 0.115\,{\scriptsize$\pm$0.007} \\
NCEM-GCN & 0.130\,{\scriptsize$\pm$0.002} & 0.358\,{\scriptsize$\pm$0.001} & 0.028\,{\scriptsize$\pm$0.000} & 0.171\,{\scriptsize$\pm$0.001} & 0.056\,{\scriptsize$\pm$0.004} & 0.236\,{\scriptsize$\pm$0.001} \\
GAT & 0.111\,{\scriptsize$\pm$0.001} & 0.327\,{\scriptsize$\pm$0.000} & 0.016\,{\scriptsize$\pm$0.002} & 0.128\,{\scriptsize$\pm$0.000} & 0.043\,{\scriptsize$\pm$0.003} & 0.201\,{\scriptsize$\pm$0.006} \\
$k$-NN prop.\ ($k{=}49$) & 0.076\,{\scriptsize$\pm$0.004} & 0.268\,{\scriptsize$\pm$0.007} & 0.026\,{\scriptsize$\pm$0.002} & 0.159\,{\scriptsize$\pm$0.007} & 0.065\,{\scriptsize$\pm$0.002} & 0.250\,{\scriptsize$\pm$0.002} \\
$k$-NN prop.\ ($k{=}25$) & 0.074\,{\scriptsize$\pm$0.003} & 0.264\,{\scriptsize$\pm$0.005} & 0.023\,{\scriptsize$\pm$0.001} & 0.151\,{\scriptsize$\pm$0.002} & 0.061\,{\scriptsize$\pm$0.000} & 0.241\,{\scriptsize$\pm$0.000} \\
$k$-NN prop.\ ($k{=}10$) & 0.072\,{\scriptsize$\pm$0.001} & 0.258\,{\scriptsize$\pm$0.002} & 0.019\,{\scriptsize$\pm$0.000} & 0.137\,{\scriptsize$\pm$0.001} & 0.049\,{\scriptsize$\pm$0.002} & 0.217\,{\scriptsize$\pm$0.003} \\
\bottomrule
\end{tabular}
\end{table}

\begin{table}[t]
\centering
\caption{Ablations on the Xenium Breast held-out sample at matched training steps. Variance explained on the measured panel (s.d.\ over five seeds $\le 0.0054$ throughout); $\Delta$ is the relative change against the full model, $R^2=0.1684$.}
\label{tab:ablation}
\footnotesize
\setlength{\tabcolsep}{4pt}
\begin{tabular}{lrr@{\hskip 16pt}lrr}
\toprule
Setting & $R^2$ & $\Delta$ & Setting & $R^2$ & $\Delta$ \\
\midrule
\emph{Training cells} & &  & \emph{Neighborhood size} & &  \\
\quad $2{,}347{,}589$ & $0.1290$ & -23\% & \quad $k{=}49$ (full) & $0.1684$ &  \\
\quad $3{,}043{,}399$ & $0.1467$ & -13\% & \quad $k{=}35$ & $0.1603$ & -5\% \\
\quad $4{,}254{,}069$ (full) & $0.1684$ &  & \quad $k{=}25$ & $0.1398$ & -17\% \\
\emph{Model size} & &  & \quad $k{=}10$ & $0.1022$ & -39\% \\
\quad $2$B (full) & $0.1684$ &  & \emph{Input components} & &  \\
\quad width$/2$ ($0.9$B) & $0.1577$ & -6\% & \quad full model & $0.1684$ &  \\
\quad width$/4$ ($0.4$B) & $0.1489$ & -12\% & \quad no direct-contact emb. & $0.1529$ & -9\% \\
 & &  & \quad no distance emb. & $0.1236$ & -27\% \\
\bottomrule
\end{tabular}
\end{table}

\subsection{Ablation studies on model components and hyperparameters}
\label{sec:scaling}

Four ablations, all evaluated on the held-out breast cancer dataset also used in the benchmarking section, were performed to evaluate whether data scaling, model scaling, model hyperparameter changes, and architecture ablations would lead to decreased model performance (Table~\ref{tab:ablation}). More training cells improve variance explained monotonically from
$0.1290$ at $2.35$M cells to $0.1684$ at $4.25$M. Since the steps are matched, the gain is
attributable to corpus size and diversity rather than to additional optimization. Halving and
quartering the model width degrades its performance from $0.1684$ at $2$B parameters to $0.1577$ and
$0.1489$, showing the model scales with capacity. Shrinking the neighborhood is the most damaging
ablation of the four: $k=35$, $25$, and $10$ give $0.1603$, $0.1398$, and $0.1022$, the last a
$39\%$ relative loss, indicating that the signal extends well beyond the immediately adjacent
cells. Removing the geometry inputs is the next most damaging: dropping the direct-contact
embedding gives $0.1529$, and dropping the distance embedding gives $0.1236$, a $27\%$ relative
loss. This confirms that spatial information is crucial to the cell state--niche correlation. In addition, we trained the GITIII-scale model purely on the three evaluation datasets with the same cross-validation and found that the model cannot converge on a single dataset. As an ablation study, we also trained GITIII on the three evaluation datasets and found that dataset-specific training leads to about $4\%$ to $9\%$ decrease in variance explained across the three datasets compared to GITIII-scale (Table~\ref{tab:datasetspecific}).

The two prediction heads also let us quantify the cost of interpretability directly. Across the same runs,
the additive head~1, whose predictions decompose exactly into the influence tensor, reaches an
average gene PCC within $2\%$ of the graph-transformer head~2 ($0.1902$ vs.\ $0.1934$ at the
selected checkpoint; $0.3702$ vs.\ $0.3726$ on the training corpus). Exact decomposability
therefore costs almost nothing in accuracy.

\subsection{Case study: an interpretable endothelial proliferation axis}
\label{sec:casestudy}

\begin{figure}[t]
\centering
\includegraphics[width=\textwidth]{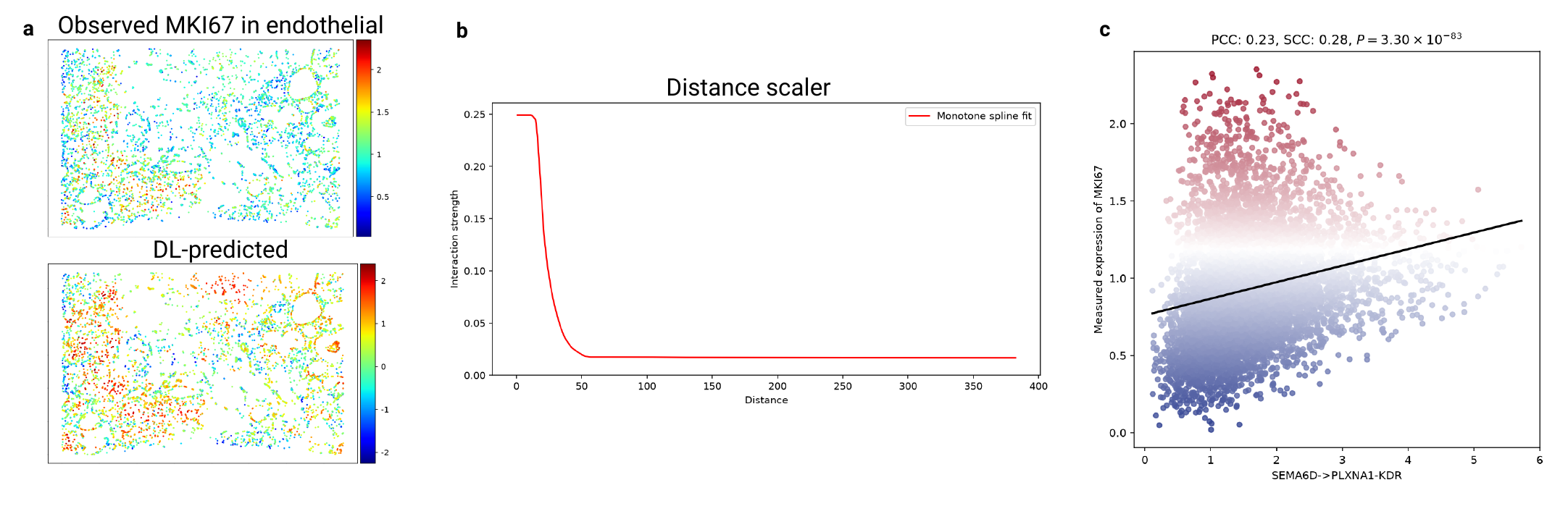}
\caption{Interpretability case study in the held-out breast sample.
\textbf{(a)} Observed and predicted \emph{MKI67} in endothelial cells.
\textbf{(b)} Fitted distance scaler for this (receiver type, target gene) pair.
\textbf{(c)} Received KDR-axis signal strength versus measured \emph{MKI67} in receiver
endothelial cells.}
\label{fig:case}
\end{figure}

To demonstrate our model's interpretability, we asked which signaling axis is most correlated with expression of \emph{MKI67}
(Ki-67), a proliferation marker whose expression level is routinely used for tumor grading and
prognosis \citep{davey2021ki67}. The
model reproduces the spatial pattern of \emph{MKI67} in this cell type
(Figure~\ref{fig:case}a) with similar spatial expression patterns between the observed and predicted. The fitted
distance scaler (Figure~\ref{fig:case}b) decays sharply within $\sim$20--50\,$\mu$m
and then plateaus, aligning with expected ligand diffusion physical law. Applying the
attribution procedure of Section~\ref{sec:attribution} selects the KDR (VEGFR2) axis as the
top-ranked pathway, and the received signal strength correlates with measured \emph{MKI67} in
receiver endothelial cells (Figure~\ref{fig:case}c; PCC $0.23$, $P = 3.3 \times 10^{-83}$,
$n = 7{,}064$). Repeating the selection five times returns the same top-ranked axis each time.
VEGFR-related signaling pathway is the canonical driver of endothelial proliferation and the
target of anti-angiogenic therapy \citep{ferrara2003vegf,alsanea2021vegfr2}, so the
model recovers a well-established mechanism in a tumor type absent from its training corpus.

\section{Conclusion}
\label{sec:conclusion}

We present GITIII-scale, an interpretable graph foundation model for cell state--niche correlation analysis in spatial transcriptomics. GITIII-scale uses transformers to model the interaction between two single cells at a given spatial distance and a graph transformer to model the cellular neighborhood features. Trained on our assembled pan-cancer, specimen-level matched scRNA-seq and IST database with 4,254,069 cells, GITIII-scale overcomes previous databases' drawback of losing either single-cell resolution or full transcriptomics measurement, and can produce TME embeddings that capture cell state--niche correlation better than previous foundation models. Its unique interpretable architecture also enables the single-cell-level analysis of how each gene of each receiver cell is influenced by each of its neighboring sender cell, and further statistical analysis of which pathway is highly correlated with the target gene expression in a given receiver cell type, enabling an interpretable modelling of intercellular interaction, the TME, and drug target discovery studies.

The limitations of this study, together with the future directions they motivate, are discussed
in Appendix~\ref{sup:limitations}.

\subsubsection*{Ethics Statement}

All datasets used in this study are publicly available and were obtained from previously
published works with appropriate citations. These datasets were originally collected under IRB
approval or equivalent ethical review, with informed consent obtained as documented in the
source publications. We only use fully anonymized data and did not conduct any new data
collection or direct interactions with human subjects. No personally identifiable information or
sensitive patient data is included. Our methods are designed to advance scientific understanding
of spatial transcriptomics and biological modeling and do not pose foreseeable risks related to
privacy, misuse, or social harm.

\subsubsection*{Declaration of Large Language Models Usage in This Work}

Large Language Models (LLMs) were used as a writing aid to enhance clarity and readability
during the preparation of this manuscript. The use of these tools was strictly limited to grammatical
correction and stylistic refinements. LLMs (Claude and ChatGPT) were additionally used to create
or modify the scientific figures and images in this manuscript. All intellectual
content, analyses, and arguments in the paper did not rely on the use of LLMs in any way.

\subsubsection*{Reproducibility Statement}

All datasets used are public; their platforms, panel sizes, and cell counts are listed in
Table~\ref{tab:accessions} of the supplementary material. The original data of the pretraining data used in this paper can be found in their corresponding papers. The imputed pretraining data, codes and the model checkpoint to reproduce the results are available at \url{https://github.com/lugia-xiao/GITIII_scale}.

\subsubsection*{Acknowledgements}

This study is personally funded by Marissa Mao.

\clearpage
\renewcommand{\refname}{References}

\clearpage
\setcounter{section}{0}
\setcounter{table}{0}
\setcounter{figure}{0}
\setcounter{equation}{0}

\section*{Supplementary Material}
\addcontentsline{toc}{section}{Supplementary Material}

\renewcommand{\thesection}{S\arabic{section}}
\renewcommand{\thetable}{S\arabic{table}}
\renewcommand{\theequation}{S\arabic{equation}}
\renewcommand{\theHsection}{S\arabic{section}}
\renewcommand{\theHsubsection}{S\arabic{section}.\arabic{subsection}}
\renewcommand{\theHtable}{S\arabic{table}}
\renewcommand{\theHequation}{S\arabic{equation}}
\renewcommand{\theHfigure}{S\arabic{figure}}

\section{Limitations and future directions}
\label{sup:limitations}

Our study has three limitations, which set our future directions. 1. Specimen-level matched
scRNA-seq and IST data are rare, so we will keep collecting data to scale the model further;
limited compute also meant imputation was performed with SpaIM alone. 2. LR attribution accuracy
is hard to evaluate systematically without ground truth, and our analysis establishes correlation
rather than causality, so further biological validation is needed. Similarly, exact decomposition in the interpretability analysis is a mathematical association estimation and does not establish uniquely biologically meaningful sender cell's influences, where real sender effects still require wet-lab experiments. 3. Biological interpretation
requires inputs imputed from specimen-level paired IST and scRNA-seq data, which limits the
datasets our model can be applied to.

Finally, the benchmarked foundation models were not designed to capture cell state--niche
relationships, and their original input formulations may not involve predicting masked expression
from the niche. The comparison here therefore reflects differences in modeling objectives rather
than model quality.

\section{Benchmark under a smaller probe}
\label{sup:benchtables}

Table~\ref{tab:bench256} repeats the benchmark of Table~\ref{tab:bench} with a $256$-unit probe
hidden layer instead of $1024$. Every entry uses the masking convention, data split, and
training recipe of Section~\ref{sup:extraction}, so the two differ only in the width of the
probe's hidden layer.

\begin{table}[!ht]
\centering
\caption{The benchmark of Table~\ref{tab:bench} repeated with a $256$-unit probe hidden layer, all else identical. The ranking is unchanged, so the margin of GITIII-scale is not an artifact of probe capacity.}
\label{tab:bench256}
\small
\setlength{\tabcolsep}{3pt}
\begin{tabular}{l cc cc cc}
\toprule
 & \multicolumn{2}{c}{Xenium Breast} & \multicolumn{2}{c}{MERFISH CRC} & \multicolumn{2}{c}{MERFISH Melanoma} \\
\cmidrule(lr){2-3}\cmidrule(lr){4-5}\cmidrule(lr){6-7}
Method & $R^2$ & PCC & $R^2$ & PCC & $R^2$ & PCC \\
\midrule
\multicolumn{7}{l}{\emph{Measured genes}} \\
\textbf{GITIII-scale (ours)} & \textbf{0.170}\,{\scriptsize$\pm$0.004} & \textbf{0.400}\,{\scriptsize$\pm$0.005} & \textbf{0.064}\,{\scriptsize$\pm$0.003} & \textbf{0.251}\,{\scriptsize$\pm$0.006} & \textbf{0.106}\,{\scriptsize$\pm$0.006} & \textbf{0.320}\,{\scriptsize$\pm$0.010} \\
TERRA & 0.145\,{\scriptsize$\pm$0.001} & 0.369\,{\scriptsize$\pm$0.002} & 0.036\,{\scriptsize$\pm$0.002} & 0.194\,{\scriptsize$\pm$0.004} & 0.076\,{\scriptsize$\pm$0.005} & 0.275\,{\scriptsize$\pm$0.009} \\
HEIST & 0.121\,{\scriptsize$\pm$0.000} & 0.340\,{\scriptsize$\pm$0.001} & 0.043\,{\scriptsize$\pm$0.001} & 0.207\,{\scriptsize$\pm$0.002} & 0.077\,{\scriptsize$\pm$0.002} & 0.275\,{\scriptsize$\pm$0.003} \\
Novae & 0.111\,{\scriptsize$\pm$0.002} & 0.324\,{\scriptsize$\pm$0.003} & 0.029\,{\scriptsize$\pm$0.001} & 0.168\,{\scriptsize$\pm$0.002} & 0.044\,{\scriptsize$\pm$0.001} & 0.206\,{\scriptsize$\pm$0.003} \\
SpatialFormer & 0.057\,{\scriptsize$\pm$0.010} & 0.223\,{\scriptsize$\pm$0.020} & 0.012\,{\scriptsize$\pm$0.003} & 0.121\,{\scriptsize$\pm$0.002} & 0.014\,{\scriptsize$\pm$0.005} & 0.106\,{\scriptsize$\pm$0.016} \\
NCEM-GCN & 0.128\,{\scriptsize$\pm$0.006} & 0.352\,{\scriptsize$\pm$0.005} & 0.021\,{\scriptsize$\pm$0.007} & 0.149\,{\scriptsize$\pm$0.018} & 0.052\,{\scriptsize$\pm$0.003} & 0.227\,{\scriptsize$\pm$0.009} \\
GAT & 0.102\,{\scriptsize$\pm$0.006} & 0.311\,{\scriptsize$\pm$0.006} & 0.015\,{\scriptsize$\pm$0.002} & 0.132\,{\scriptsize$\pm$0.002} & 0.032\,{\scriptsize$\pm$0.003} & 0.185\,{\scriptsize$\pm$0.012} \\
$k$-NN prop.\ ($k{=}49$) & 0.080\,{\scriptsize$\pm$0.002} & 0.269\,{\scriptsize$\pm$0.004} & 0.028\,{\scriptsize$\pm$0.001} & 0.159\,{\scriptsize$\pm$0.001} & 0.061\,{\scriptsize$\pm$0.005} & 0.237\,{\scriptsize$\pm$0.009} \\
$k$-NN prop.\ ($k{=}25$) & 0.081\,{\scriptsize$\pm$0.002} & 0.269\,{\scriptsize$\pm$0.004} & 0.021\,{\scriptsize$\pm$0.001} & 0.135\,{\scriptsize$\pm$0.004} & 0.053\,{\scriptsize$\pm$0.001} & 0.220\,{\scriptsize$\pm$0.003} \\
$k$-NN prop.\ ($k{=}10$) & 0.081\,{\scriptsize$\pm$0.001} & 0.269\,{\scriptsize$\pm$0.001} & 0.018\,{\scriptsize$\pm$0.001} & 0.123\,{\scriptsize$\pm$0.003} & 0.045\,{\scriptsize$\pm$0.000} & 0.203\,{\scriptsize$\pm$0.001} \\
\midrule
\multicolumn{7}{l}{\emph{All genes}} \\
\textbf{GITIII-scale (ours)} & \textbf{0.193}\,{\scriptsize$\pm$0.001} & \textbf{0.433}\,{\scriptsize$\pm$0.001} & \textbf{0.069}\,{\scriptsize$\pm$0.002} & \textbf{0.262}\,{\scriptsize$\pm$0.005} & \textbf{0.119}\,{\scriptsize$\pm$0.002} & \textbf{0.337}\,{\scriptsize$\pm$0.005} \\
TERRA & 0.154\,{\scriptsize$\pm$0.002} & 0.386\,{\scriptsize$\pm$0.002} & 0.034\,{\scriptsize$\pm$0.003} & 0.193\,{\scriptsize$\pm$0.001} & 0.075\,{\scriptsize$\pm$0.003} & 0.267\,{\scriptsize$\pm$0.006} \\
HEIST & 0.124\,{\scriptsize$\pm$0.003} & 0.348\,{\scriptsize$\pm$0.004} & 0.041\,{\scriptsize$\pm$0.002} & 0.204\,{\scriptsize$\pm$0.006} & 0.072\,{\scriptsize$\pm$0.001} & 0.268\,{\scriptsize$\pm$0.002} \\
Novae & 0.117\,{\scriptsize$\pm$0.002} & 0.337\,{\scriptsize$\pm$0.003} & 0.031\,{\scriptsize$\pm$0.001} & 0.175\,{\scriptsize$\pm$0.003} & 0.050\,{\scriptsize$\pm$0.002} & 0.220\,{\scriptsize$\pm$0.003} \\
SpatialFormer & 0.047\,{\scriptsize$\pm$0.000} & 0.202\,{\scriptsize$\pm$0.001} & 0.017\,{\scriptsize$\pm$0.002} & 0.078\,{\scriptsize$\pm$0.009} & 0.019\,{\scriptsize$\pm$0.001} & 0.130\,{\scriptsize$\pm$0.003} \\
NCEM-GCN & 0.123\,{\scriptsize$\pm$0.005} & 0.347\,{\scriptsize$\pm$0.008} & 0.026\,{\scriptsize$\pm$0.004} & 0.167\,{\scriptsize$\pm$0.003} & 0.053\,{\scriptsize$\pm$0.008} & 0.225\,{\scriptsize$\pm$0.018} \\
GAT & 0.103\,{\scriptsize$\pm$0.015} & 0.318\,{\scriptsize$\pm$0.020} & 0.015\,{\scriptsize$\pm$0.004} & 0.122\,{\scriptsize$\pm$0.019} & 0.040\,{\scriptsize$\pm$0.001} & 0.196\,{\scriptsize$\pm$0.004} \\
$k$-NN prop.\ ($k{=}49$) & 0.072\,{\scriptsize$\pm$0.001} & 0.260\,{\scriptsize$\pm$0.002} & 0.022\,{\scriptsize$\pm$0.010} & 0.141\,{\scriptsize$\pm$0.037} & 0.064\,{\scriptsize$\pm$0.000} & 0.245\,{\scriptsize$\pm$0.001} \\
$k$-NN prop.\ ($k{=}25$) & 0.074\,{\scriptsize$\pm$0.000} & 0.264\,{\scriptsize$\pm$0.000} & 0.015\,{\scriptsize$\pm$0.003} & 0.120\,{\scriptsize$\pm$0.010} & 0.063\,{\scriptsize$\pm$0.003} & 0.246\,{\scriptsize$\pm$0.007} \\
$k$-NN prop.\ ($k{=}10$) & 0.074\,{\scriptsize$\pm$0.002} & 0.261\,{\scriptsize$\pm$0.003} & 0.018\,{\scriptsize$\pm$0.000} & 0.129\,{\scriptsize$\pm$0.003} & 0.050\,{\scriptsize$\pm$0.001} & 0.220\,{\scriptsize$\pm$0.002} \\
\bottomrule
\end{tabular}
\end{table}

\section{Related work}
\label{sup:related}

\paragraph{Foundation models for spatial transcriptomics.}
Individual spatial transcriptomics slides are small, noisy, and measured on incompatible
gene panels, so models fit to one slide neither generalize nor pool information across
studies. Spatial foundation models are designed to overcome this. Such a model is pretrained once on
many slides with a self-supervised objective, and then exposes a frozen per-cell or per-niche
embedding. Downstream users probe that embedding with a light task head instead of training a new
model for each dataset. scGPT-spatial \citep{wang2025scgptspatial} continually pretrains a gene-token transformer from spot- and cell-resolution slides. Its objective masks a subset of a
cell's gene tokens and regresses their binned expression values, with a neighborhood term
that asks the same decoder to reconstruct a spot from the embeddings of its spatially
adjacent spots. It outputs contextual cell and gene embeddings, and is applied to
cell-type deconvolution of spots and to imputation of unmeasured genes.
Nicheformer \citep{schaar2025nicheformer} tokenizes each cell as a gene sequence ordered by
expression rank and trains a transformer encoder with a masked-language-model objective,
predicting the identity of randomly masked gene tokens by cross-entropy over the gene
vocabulary. It outputs a per-cell embedding, which is linearly
probed or fine-tuned to predict spatial context labels such as niche and region for
dissociated cells. Because that embedding represents the cell itself rather than its
surrounding niche (the model consumes one cell's gene sequence and never a neighborhood),
Nicheformer produces no niche-level representation to supply to our probe, and we therefore
exclude it from the benchmark. Novae \citep{blampey2025novae} encodes a cell's
neighborhood subgraph with a graph attention network. Rather than reconstructing
expression, it clusters neighborhoods into a set of learned prototypes and trains the
encoder so that two randomly perturbed versions of the same neighborhood are assigned to
the same prototype. It outputs a per-cell
niche embedding and a discrete niche assignment used for spatial domain annotation. HEIST \citep{madhu2025heist} models a tissue as
a hierarchical graph, a spatial cell graph whose nodes each carry a gene co-expression
network, and pretrains a hierarchical graph transformer on IST datasets. It outputs
gene-level embeddings and niche embeddings that are used for cell typing, gene imputation,
and clinical outcome prediction. SpatialFormer \citep{wang2026spatialformer} combines
convolutional and transformer blocks over rank-ordered gene tokens and is pretrained on cell
pairs with masked-token, subcellular transcript-adjacency, and pairwise-proximity objectives.
It outputs cell embeddings applied to batch correction, cell-type annotation, and
co-localization detection. TERRA \citep{birk2026terra} tokenizes an index cell together with
its neighborhood as one gene-token sequence and pretrains a joint-embedding predictive
architecture that predicts masked molecular and spatial context in latent space. It outputs
gene, cell, and neighborhood embeddings from one backbone, used zero-shot for niche and
cell-type identification and for in silico gene perturbation.
 These models are designed for
representing cells and their spatial context, and are not intended for cell state--niche
correlation analysis.

\paragraph{Models and tools for cell--cell interaction analysis.}
Most previous CCI methods focus on estimating which interaction events occur between which
two cell types and through which LR pair. The candidate LR pairs are
supplied by established databases, and each candidate is scored by a statistical test on
co-expression: permutation of cell-type labels in CellPhoneDB
\citep{efremova2020cellphonedb,troule2025cellphonedbv5}, a mass-action score with cofactor
terms in CellChat \citep{jin2021cellchat}, and consensus scoring across resources in LIANA+
\citep{dimitrov2024liana}. Later tools for spatial transcriptomics incorporate spatial
information into this scoring, either with graph neural networks over the cell graph or by
distance-based thresholding and weighting of candidate interactions
\citep{cang2023commot,li2023spatialdm,palla2022squidpy}. Only a few methods instead
estimate the downstream effect of a CCI on the receiver cell: NicheNet
\citep{browaeys2020nichenet} propagates ligands through a prior signaling and
gene-regulatory network to a ligand--target regulatory potential matrix. It inherits the coverage of
the prior they are built on, so signaling and target genes that are not yet curated fall
outside the hypothesis space regardless of the evidence a dataset contains. NCEM
\citep{fischer2023ncem} regresses a cell's expression on its neighborhood composition with
graph convolutional models, and GITIII \citep{xiao2025gitiii} together with graph-attention
models such as HoloNet \citep{li2023holonet} learns dataset specific cell state-niche correlation and attribution that correlation to certain cells or pathways.
The data-driven models
estimate the relationship from the data instead, but are typically fitted to one dataset at
a time, so what they learn in one tissue is not carried over to the next. Estimating
neighbor-attributable state changes in a way that is both data-driven and transferable
across samples therefore remains open.

\section{Differences from GITIII}
\label{sup:vsgitiii}

GITIII-scale shares its motivating question with GITIII \citep{xiao2025gitiii}, namely
decomposing a receiver cell's state into contributions from individual neighboring senders, but
differs from it in three respects that together are what make a pretrained, transferable model possible.

\paragraph{Architecture.} GITIII models a neighborhood in a single pass over the sender set. In
GITIII-scale the sender--receiver relationship is instead built in two levels. A pairwise
transformer first encodes each (sender, receiver, distance, contact) tuple independently. The
sending and receiving roles have separate encoders, and intercellular distance is carried as its
own token rather than as a scalar weight. Cross-attention pooling then compresses each pair to a
fixed-width vector. The two prediction heads then act on these pair
vectors. The first is a single-layer graph transformer with no feed-forward block; its per-gene
attention is normalized across senders, so its prediction is an exact sum of per-sender,
per-gene influences. That property is what the interpretability analysis depends on. The second
head is a deeper stack, which trades exact decomposability for accuracy and supplies the
cellular-neighborhood embedding used as the TME representation. Because one model yields both an
exactly decomposable influence tensor and a niche-level embedding, GITIII-scale serves as a
foundation model and an attribution tool simultaneously rather than as one or the other.

\paragraph{Pan-cancer pretraining rather than per-dataset fitting.} GITIII is fitted to a single
dataset at a time, so nothing it learns in one tissue transfers to the next, and each new sample
requires training from scratch. GITIII-scale is pretrained once across a pan-cancer corpus of
$4{,}254{,}069$ cells spanning several tumor contexts and two imaging platforms, and is then
applied frozen to held-out cancer types and platforms. Section~\ref{sup:datasetspecific}
quantifies the benefit. Dataset-specific GITIII training loses variance explained on all three
evaluation samples relative to the pretrained model, and GITIII-scale trained from scratch on a
single evaluation dataset does not converge at all.

\paragraph{Spatial single-cell data with full-transcriptome coverage.} GITIII operates on the
measured panel of an imaging-based platform, a few hundred to a few thousand genes, so the
influence it estimates is defined only over that panel. Every sample in the GITIII-scale corpus
is instead paired with scRNA-seq from the same specimen, which allows the measured panel to be
imputed onto a shared $20{,}000$-gene vocabulary. Three consequences follow. The influence
tensor is defined over a transcriptome-scale gene set rather than a panel. A single vocabulary is
shared across platforms whose panels are otherwise incompatible, without which one pretrained
model could not span them. And ligand--receptor attribution can reach axes whose genes appear on
no single platform's panel.

\section{Hyperparameters and architecture specification}
\label{sup:hparams}

Table~\ref{tab:hparams} lists every architecture and optimization hyperparameter of the
GITIII-scale run reported in the main text. At this configuration, GITIII-scale has $2{,}296{,}747{,}232$ ($\sim$2.3B) parameters
and is pretrained on the $4{,}254{,}069$ spatial cells of Table~\ref{tab:accessions}: $10$
samples spanning $4$ tumor contexts and $2$ imaging platforms. We use AdamW at a constant learning rate of
$10^{-4}$ and batch size $344$ for $2$ epochs on a single NVIDIA B200 GPU, reshuffling
dataset order each epoch, and select the checkpoint with the best macro-averaged validation
correlation. The model is trained for 23 hours.

\begin{table}[!ht]
\centering
\caption{GITIII-scale architecture and optimization hyperparameters.}
\label{tab:hparams}
\small
\setlength{\tabcolsep}{4pt}
\begin{tabular}{llc}
\toprule
Group & Hyperparameter & Value \\
\midrule
Input & Gene vocabulary $G$ & $20{,}000$ \\
      & Neighborhood size $k$ (senders per receiver) & $49$ \\
      & Cells per training example ($1$ receiver $+$ $k$ senders) & $50$ \\
      & Input tensor shape (batch, cells, genes) & $(344, 50, 20{,}000)$ \\
\midrule
Tokenization & Tokens per cell $T$ & $32$ \\
      & Token width / embedding dimension $d$ & $256$ \\
      & Attention heads & $16$ \\
      & MLP expansion ratio & $4.0$ \\
      & Distance features per sender--receiver pair & $5$ \\
      & Contact token each & $1$ \\
\midrule
Pair encoder & Sender self-attention blocks (SAB) & $3$ \\
      & Interaction self-attention blocks (SAB) & $3$ \\
      & Pair-vector dimension \texttt{out\_edge\_dim} & $256$ \\
      & MCAB latent queries, pair encoder & $4$ \\
\midrule
Head 1 (additive) & Per-gene attention normalization & across senders \\
      & Output & influence tensor $\mathcal{I}_i \in \R^{k \times G}$ \\
\midrule
Head 2 (niche) & Transformer blocks \texttt{num\_layers\_deeper} & $4$ \\
      & MCAB latent queries $L$ & $4$ \\
      & TME embedding dimension $L \cdot d$ & $1024$ \\
\midrule
Regularization & Dropout & $0.0$ \\
      & Attention dropout & $0.0$ \\
\midrule
Objective & Per-gene loss & $1 - \mathrm{PCC}$ across cells \\
      & Gene weighting $w^{(g)}$ & $\propto \mathrm{std}_i(z_i^{(g)})$, mean-norm. \\
      & Magnitude term weight $\lambda$ & $0.05$ \\
      & Numerical guard $\epsilon$ & $10^{-8}$ \\
      & Stratification & receiver cell type, macro-avg. \\
      & Heads optimized & both, losses summed \\
\midrule
Optimization & Optimizer & AdamW \\
      & Learning rate & $10^{-4}$ (constant) \\
      & Batch size & $344$ \\
      & Epochs & $2$ \\
      & Dataset order & reshuffled each epoch \\
      & Checkpoint selection & best validation PCC \\
\midrule
Hardware & GPU & $1 \times$ NVIDIA B200 \\
      & Pretraining wall-clock & $23$\,h \\
\bottomrule
\end{tabular}
\end{table}

\paragraph{Cell-type-aware input construction.} A sender cell's input vector is its own
measured cell state expression added to its cell-type mean profile, except where the sender shares the
receiver's cell type, in which case only the cell-type mean is supplied. This is the same-type
masking of Section~\ref{sec:arch} of the main text, applied at the input rather than as a separate
preprocessing pass, and it removes the receiver's own state and the states of same-type
neighbors from every training example.

\paragraph{Pooling.} Both levels pool with a multi-head cross-attention block (MCAB) in which
learned latent queries, initialized as $\mathcal{N}(0, 0.02^2)$, attend to the token set and
return a fixed-size output independent of the number of input tokens. The pair vector and the TME
embedding are therefore permutation-invariant with respect to sender ordering.

\section{Distance featurization}
\label{sup:distance}

The Euclidean separation $d_{ij}$ between receiver $i$ and sender $j$, in microns, is expanded
into five scalar features before being mapped to the distance token,
\begin{equation}
\phi(d) = \left[\;
\frac{10}{d+1},\quad
\frac{10}{\sqrt{d}+1},\quad
\frac{100}{d^{3}+1},\quad
\frac{10}{d+1}e^{-d/0.76},\quad
\frac{10}{d+1}e^{-d/190.67}
\;\right] .
\label{eq:distfeat}
\end{equation}
The five terms span the decay regimes that intercellular signaling is expected to occupy. Two
are slow algebraic decays and one is a steep cubic decay that acts as a contact-range indicator.
The remaining two are exponentially damped, with length constants of $0.76$\,$\mu$m and
$190.67$\,$\mu$m that bracket juxtacrine and diffusive transport. Separate linear maps embed $\phi(d_{ij})$ for the
sender stage and for the interaction stage, so the two stages can weight the regimes
differently. The model is given no distance prior beyond Equation~\ref{eq:distfeat}; the
effective interaction range reported in the case study of the main text is fit from data.

A binary contact token, taken from whether $i$ and $j$ share an edge in the Delaunay
triangulation, is embedded alongside the distance token.

\section{Per-model embedding-extraction protocol}
\label{sup:extraction}

Every method is evaluated under the identical masking convention, data split, downstream probe,
and training recipe described in Section~\ref{sec:exp-metrics} of the main text; the only variable is the
representation handed to the probe. Methods divide into two groups. The four pretrained spatial
foundation models (Novae, HEIST, TERRA, SpatialFormer) are used as \emph{frozen} encoders. Each is
run through its own authors' released preprocessing and inference path rather than a
reimplementation, so that the pretrained weights receive the input distribution and graph
topology they were trained on. The resulting per-cell embedding is computed once per dataset and
reused unchanged across every probe configuration and replicate. The remaining baselines
(NCEM-GCN, GAT, and the $k$-NN cell-type-proportion probe) have no pretrained component and are
retrained end to end for every configuration and replicate. Table~\ref{tab:extraction}
summarizes the resulting representations.

\begin{table}[!ht]
\centering
\caption{Representations supplied to the shared probe. All methods see the same masked
neighborhoods of the same cells and the same $0.7/0.15/0.15$ split. $D$ is the dimensionality of
the per-cell representation reaching the probe.}
\label{tab:extraction}
\small
\setlength{\tabcolsep}{3pt}
\begin{tabular}{llc}
\toprule
Method & Representation & $D$ \\
\midrule
GITIII-scale  & TME embedding $\vh_i$ (head 2)              & $1024$ \\
Novae         & niche representation $z$                     & $64$   \\
HEIST         & cell $\oplus$ mean-pooled gene-network       & $256$  \\
TERRA         & mean-pooled context tokens                   & $384$  \\
SpatialFormer & mean CLS state over center--neighbor pairs    & $512$  \\
NCEM-GCN      & center node after 2 GCNConv layers           & $256$  \\
GAT           & center node after 2 GATConv layers           & $256$  \\
$k$-NN proportion & niche cell-type composition              & $n_{t}$ \\
\bottomrule
\end{tabular}
\end{table}

\paragraph{GITIII-scale.} The TME embedding is the pooled output of the second-level head,
$\vh_i \in \R^{1024}$, read from the checkpoint selected on validation correlation. No
fine-tuning is performed on any held-out sample.

\subsection{Pretrained foundation models: checkpoints and inference API}
\label{sup:api}

For these four methods we record only the checkpoint identifier and the exact library calls
used, since all preprocessing, graph construction, and tokenization are the authors' own and are
invoked unmodified.

\paragraph{Novae.} Checkpoint \texttt{prism-oncology/novae-human-0}
(\url{https://huggingface.co/prism-oncology/novae-human-0}), loaded with
\texttt{novae.Novae.from\_pretrained(...)} followed by \texttt{.eval()}. The spatial graph is
built by \texttt{novae.spatial\_neighbors(adata, radius=100)} and each center cell's local
subgraph uses the \texttt{n\_hops\_local} value stored in the checkpoint rather than an override.
Novae's internal \texttt{model.\_prepare\_adatas} performs gene selection against the model's own
vocabulary and \texttt{ZScoreTorch} standardizes the selected matrix. Per batch, subgraphs are
packed as \texttt{torch\_geometric.data.Data} objects, collated with
\texttt{Batch.from\_data\_list}, and embedded under \texttt{torch.no\_grad()} as
\texttt{model.encoder(model.cell\_embedder(batch))}. The embedding is that encoder output,
$D=\texttt{model.hparams.output\_size}=64$.

\paragraph{HEIST.} Checkpoint \texttt{HirenMadhu/HEIST}
(\url{https://huggingface.co/HirenMadhu/HEIST}), with the \texttt{GraphEncoder} class taken from
the authors' repository (\url{https://github.com/Graph-and-Geometric-Learning/HEIST}) and loaded
via \texttt{GraphEncoder.from\_pretrained(...)} at the released configuration
(\texttt{pe\_dim} $=$ \texttt{hidden\_dim} $=$ \texttt{output\_dim} $=128$, $10$ layers, $8$
heads, cross-message-passing and positional encoding enabled), then \texttt{.eval()}. Inference
uses a custom batched encode rather than the library's stock \texttt{model.encode()}. The stock
path lets multi-head attention attend across all cells in a batch, whereas attention must be
scoped to each cell's own $50$-node neighborhood. The custom path reshapes flat
$(B\!\cdot\!L, D)$ tensors to $(B, L, D)$ for the attention call only, running the GIN
convolutions, \texttt{TransformerConv} cross-message-passing, and \texttt{LayerNorm} on
flattened block-diagonal tensors, which is numerically equivalent for those submodules. The
embedding concatenates the center node's $128$-dimensional cell-level representation with the
$128$-dimensional mean-pooled representation of its gene co-expression network, $D=256$.

\paragraph{TERRA.} Checkpoint \texttt{Lotfollahi-lab/TERRA-96M}
(\url{https://huggingface.co/Lotfollahi-lab/TERRA-96M}). The model is constructed by the
library's \texttt{init\_model(...)} from the checkpoint's own \texttt{model\_config.yaml}
(\texttt{enc\_emb\_dim}$=384$, \texttt{enc\_depth}$=12$, \texttt{num\_heads}$=6$), and weights are
loaded into the target-encoder branch (the JEPA context-representation branch used for
embeddings, not the online encoder or predictor) via
\texttt{load\_checkpoint(..., target\_encoder=..., is\_training=False)}, then \texttt{.eval()}. Number of neighbors selection follows the method's original codebase's default option. The spatial graph is TERRA's own \texttt{construct\_neighbor\_graph} with
\texttt{n\_neighs=10}, matching its pretraining configuration. Inference calls
\texttt{target\_encoder.backbone.return\_layer\_emb(layers=[12], ..., ignore\_spc\_tokens=True)}
under \texttt{torch.inference\_mode()} with \texttt{bfloat16} autocast, attending only to
non-pad tokens; the per-token outputs are mean-pooled over all non-pad positions of all segments, $D=384$. 

\paragraph{SpatialFormer.} Checkpoint \texttt{ckp\_pair\_13tissues\_71.ckpt}, the variant
recommended by the authors, obtained from the Figshare asset linked in their repository
(\url{https://figshare.com/articles/dataset/pair_input_checkpoint_5k/31146247}; code at
\url{https://github.com/TerminatorJ/Spatialformer}) and checksum-verified against the published
file. Number of neighbors selection follows the method's original codebase's default option. The model is built by the repository's
\texttt{manual\_train\_fm(config=..., use\_flash\_attn=False)} from its large paired training
config with the packaged gene-embedding table, weights loaded with
\texttt{model.load\_state\_dict(...)}, then \texttt{model.eval()} and
\texttt{model.input\_type = "pair"}. Batches are collated with the library's own
\texttt{collate\_fn} to preserve its ALiBi-compatible padding layout, and embeddings are read
from \texttt{model.get\_embeddings(batch, layers=[-1], pair\_prediction=True)} under
\texttt{bfloat16} autocast, taking the CLS-position hidden state. 

\paragraph{Same cell type expression masking.} The masking rule is identical for every
method, but the representation it acts on differs by necessity, since it must be applied at
whichever stage each pretrained model accepts input. For Novae this is after its per-slide
$z$-scoring; for HEIST, on its denoised gene panel; and for TERRA and SpatialFormer, at the token
level, where a type-mean tokenization is substituted for a cell's own. The latter two require one
further departure. Their public tokenizers cache one tokenization per cell and reuse it wherever
that cell appears as another cell's neighbor, which masking cannot accommodate when the
substitution depends on the query cell's type. For those two methods we therefore reimplement the
ranking and truncation logic at the script level, verified to reproduce the library tokenizer
exactly on unmasked input.

\subsection{Baselines trained end to end}
\label{sup:e2e}

\paragraph{Shared graph for NCEM-GCN and GAT.} Both use a Delaunay triangulation truncated to
edges of at most $100\,\mu$m, from which an induced ego-subgraph (center, its direct
neighbors, and the edges among them) is extracted per center cell and cached. Both trunks apply
two message-passing layers over this subgraph, so information reaches the center cell over paths
of up to two hops. The same cached
topology is used by both architectures, so the comparison between them is not confounded by
graph construction. Node features are the reconstructed full expression over nonzero-variance
genes, $z$-scored with train-split statistics and held fixed across configurations so that the
input information budget does not vary with the target gene set. Masking replaces the $z$-scored
feature row of the center cell and of every same-type node in its ego-subgraph with the
type-mean vector, $z$-scored under the same train-split statistics.

\paragraph{NCEM-GCN.} A linear input projection to $256$ units with ReLU, two
\texttt{GCNConv} layers at width $256$, and the center node's output concatenated with a
$128$-dimensional learned cell-type embedding before the head. The graph trunk width is fixed
across configurations; only the head width and output gene count vary.

\paragraph{GAT.} Identical in every respect to NCEM-GCN except the trunk: ELU activations in
place of ReLU, and two \texttt{GATConv} layers with $4$ attention heads (the first concatenating
heads to $256$ dimensions, the second averaging them).

\paragraph{$k$-NN cell-type proportion.} This baseline consumes no expression at all. A
\texttt{cKDTree} over spatial coordinates retrieves each cell's $k$ nearest \emph{other} cells,
and the feature vector is the cell-type composition of those neighbors: a length-$n_{t}$ vector
whose entries are the fractions of the $k$ neighbors belonging to each cell type, summing to one.
It is concatenated with the same $128$-dimensional learned cell-type embedding and passed to a
head of the shared shape. Neither a normalization layer nor input standardization is needed here,
since a composition vector is already bounded in $[0,1]$ and does not exhibit the collapse
described above. Changing $k \in \{10,25,49\}$ changes only which precomputed composition array is loaded;
architecture and hyperparameters are identical across the three variants. The same-type masking
convention does not apply here and is not invoked, since there is no expression feature to
replace; the prediction target is the unmasked cell-state residual, exactly as for every other
method.

\subsection{Shared probe and training recipe}
\label{sup:probe}

Every method's representation is consumed by a probe of the same shape,
$\mathrm{Linear}(D{+}128, h) \rightarrow \mathrm{ReLU} \rightarrow
\mathrm{Linear}(h, n_{\mathrm{genes}})$, where the $128$-dimensional cell-type embedding is
learned jointly with the probe and never part of a frozen upstream model. Training uses Adam at
learning rate $10^{-3}$, mean squared error loss, early stopping on validation loss with
patience $10$ against a maximum of $200$ epochs, and restoration of the best-validation
checkpoint before test evaluation. Batch size is $1024$ for the frozen-embedding methods and the
composition baseline, and $512$ for the two graph baselines, whose batch elements are
variable-size ego-subgraphs rather than fixed-length vectors. For the frozen-embedding methods
only the probe is retrained per configuration; for the end-to-end baselines the whole model is
retrained, since the head width and target gene set enter the joint objective.

\paragraph{Masking.} For every method that consumes expression, both the receiver cell and all
same-type neighbors have their expression replaced by the cell-type mean $\vmu_{t(i)}$ before
encoding. The substitution is always applied to reconstructed full expression, never to the
cell-state residual, because the residual is the prediction target and must not enter the input.
Masking the receiver as well as its same-type neighbors is what renders the probe leakage-safe.
Cell state is spatially autocorrelated within a type, so an encoder with access to unmasked
same-type neighbors can score well simply by interpolating between them. Such a score measures how
faithfully the encoder preserves local expression, not what it captures about cell--cell
interaction.

\section{Dataset accessions}
\label{sup:datasets}

Table~\ref{tab:accessions} gives the provenance of every sample in the corpus. Cell
counts are as loaded by the pipeline; gene counts are the measured imaging panel size before
imputation onto the shared $20{,}000$-gene vocabulary. Every imaging-based sample is paired with an
scRNA-seq reference, which is what permits the imputation described in Section~\ref{sec:corpus} of the main
text.

\begin{table}[!ht]
\centering
\caption{Pan-cancer corpus: per-sample accessions and download sources. \emph{Panel} is the
measured imaging gene panel size before imputation onto the shared $20{,}000$-gene vocabulary,
and \emph{\# types} is the number of annotated cell types in the imaging sample. The three
samples in the first block are held out from pretraining and used only for out-of-distribution
evaluation; the ten below constitute the pretraining corpus.}
\label{tab:accessions}
\scriptsize
\setlength{\tabcolsep}{3pt}
\begin{tabular}{llllrrrl}
\toprule
Sample & Tumor type & Platform & Panel & IST cells & sc cells & \# types & Source \\
\midrule
\multicolumn{8}{l}{\emph{Held out from pretraining (evaluation samples)}} \\
\texttt{breast\_xenium\_ffpe} & Breast & Xenium & $312$ & $103{,}209$ & $30{,}365$ & $18$ &
  10x Genomics portal\textsuperscript{a} \\
\texttt{ecotypes\_\_CRC2} & Colorectal (met.) & MERFISH & $500$ & $38{,}080$ & $48{,}426$ & $9$ &
  SpatialEcoTyper portal\textsuperscript{d} \\
\texttt{ecotypes\_\_Melanoma1} & Melanoma (met.) & MERFISH & $500$ & $27{,}907$ & $49{,}526$ & $9$ &
  SpatialEcoTyper portal\textsuperscript{d} \\
\midrule
\multicolumn{8}{l}{\emph{Pretraining corpus}} \\
\texttt{liver\_villages\_cosmx\_\_1} & Liver (HCC/iCCA) & CosMx & $1{,}000$ & $835{,}033$ & $33{,}495$ & $6$ &
  Zenodo 13773977\textsuperscript{c} \\
\texttt{liver\_villages\_cosmx\_\_2} & Liver (HCC/iCCA) & CosMx & $1{,}000$ & $433{,}303$ & $29{,}811$ & $6$ &
  Zenodo 13773977\textsuperscript{c} \\
\texttt{liver\_villages\_cosmx\_\_3} & Liver (HCC/iCCA) & CosMx & $1{,}000$ & $867{,}430$ & $30{,}762$ & $6$ &
  Zenodo 13773977\textsuperscript{c} \\
\texttt{liver\_villages\_cosmx\_\_4} & Liver (HCC/iCCA) & CosMx & $1{,}000$ & $211{,}823$ & $18{,}438$ & $6$ &
  Zenodo 13773977\textsuperscript{c} \\
\texttt{spatch\_coad\_cosmx} & Colon adeno. & CosMx & $6{,}175$ & $292{,}371$ & $8{,}288$ & $16$ &
  SPATCH portal\textsuperscript{b} \\
\texttt{spatch\_coad\_xenium} & Colon adeno. & Xenium & $5{,}001$ & $405{,}927$ & $8{,}288$ & $16$ &
  SPATCH portal\textsuperscript{b} \\
\texttt{spatch\_hcc\_cosmx} & Hepatocellular & CosMx & $6{,}175$ & $237{,}030$ & $12{,}226$ & $17$ &
  SPATCH portal\textsuperscript{b} \\
\texttt{spatch\_hcc\_xenium} & Hepatocellular & Xenium & $5{,}001$ & $271{,}687$ & $12{,}226$ & $17$ &
  SPATCH portal\textsuperscript{b} \\
\texttt{spatch\_ov\_cosmx} & Ovarian & CosMx & $6{,}175$ & $289{,}272$ & $9{,}229$ & $13$ &
  SPATCH portal\textsuperscript{b} \\
\texttt{spatch\_ov\_xenium} & Ovarian & Xenium & $5{,}001$ & $410{,}193$ & $9{,}229$ & $13$ &
  SPATCH portal\textsuperscript{b} \\
\midrule
\textbf{Total (pretraining)} & & 2 platforms & & \textbf{$4{,}254{,}069$} & & & \\
\bottomrule
\end{tabular}
\end{table}

\noindent
\raggedright
\textsuperscript{a}\,\citet{janesick2023xenium}; 10x Genomics Xenium human breast dataset
explorer (manual download; no stable direct URL).
\textsuperscript{b}\,\citet{ren2025spatch}; SPATCH portal,
\url{http://spatch.pku-genomics.org/}.
\textsuperscript{c}\,\citet{liu2026villages}; Zenodo record 13773977 (imaging) and GEO
accession GSE189903 (scRNA-seq).
\textsuperscript{d}\,\citet{zhang2026spatialecotypes}; SpatialEcoTyper vignette portal,
\url{https://spatialecotyper.stanford.edu/}, imaging; GEO accession GSE320042, scRNA-seq. 

\paragraph{Paired imputation.} Each imaging sample and its matched scRNA-seq reference are
integrated with SpaIM \citep{li2025spaim}, which learns a per-dataset mapping from measured-panel
expression to the full scRNA-seq gene space conditioned on Leiden clusters of the reference, and
applies it to every spatial cell. Integration over the full corpus takes five days on two GPUs.

\section{Dataset-specific training comparison}
\label{sup:datasetspecific}

Table~\ref{tab:datasetspecific} compares GITIII-scale against GITIII \citep{xiao2025gitiii}
trained separately on each evaluation dataset with the same cross-validation protocol. Both
columns report variance explained on the measured gene panel, mean $\pm$ s.d. over five seeds.
GITIII-scale was also trained from scratch on each of the three evaluation datasets alone; those
runs did not converge, so no single-dataset GITIII-scale numbers are reported.

\begin{table}[H]
\centering
\caption{Dataset-specific training versus pan-cancer pretraining. Variance explained on the
measured panel, mean $\pm$ s.d. over five seeds. GITIII is trained on each evaluation dataset
separately with cross-validation; GITIII-scale is pretrained pan-cancer with all three samples
held out.}
\label{tab:datasetspecific}
\begin{tabular}{lccc}
\toprule
Dataset & GITIII (dataset-specific) & GITIII-scale (pretrained) & Relative change \\
\midrule
Xenium Breast    & $0.1530 \pm 0.0032$ & $0.1684 \pm 0.0029$ & $-9.1\%$ \\
MERFISH CRC      & $0.0565 \pm 0.0049$ & $0.0591 \pm 0.0001$ & $-4.4\%$ \\
MERFISH Melanoma & $0.1050 \pm 0.0037$ & $0.1115 \pm 0.0018$ & $-5.8\%$ \\
\bottomrule
\end{tabular}
\end{table}

\section{Mathematical justification of the interpretable analysis (influence-tensor decomposition)}
\label{sup:interpretableproof}

\subsection*{A. Pairwise construction of sender--receiver representations}

The interpretability of the GITIII-scale influence tensor follows directly from the architecture of its level-one prediction head. The key property is that the model first represents each sender--receiver interaction separately and only then aggregates these interactions to predict the receiver-cell state.

For every sender--receiver pair $(j,i)$, GITIII-scale first constructs a pair representation
\[
\ve_{ij}
=
F_\theta
\!\left(
x_i,x_j,d_{ij},c_{ij}
\right)
\in\R^d,
\]
where $d_{ij}$ denotes the spatial distance, $c_{ij}$ denotes the direct-contact feature, and $F_\theta$ denotes the pairwise transformer/tokenization pipeline. Importantly, this representation is constructed separately for each sender--receiver pair before the level-one aggregation. In the model, the sender-cell tokens, receiver-cell tokens, distance token, and direct-contact token are processed to obtain one representation $\ve_{ij}$ for each pair $(j,i)$.

This pairwise construction preserves the identity of the neighboring sender at the stage at which its contribution is defined. Although $F_\theta$ can be a nonlinear and expressive function of the sender, receiver, and spatial features, its output remains indexed by the particular sender--receiver pair $(j,i)$.

\subsection*{B. Exact additive decomposition of the predicted cell state}

For each target gene $g$, the interpretable head maps the pair representation $\ve_{ij}$ to a gene-specific attention logit and a gene-specific value. We write
\[
a_{ij}^{(g)}
=
W_A(\ve_{ij})^{(g)},
\qquad
v_{ij}^{(g)}
=
W_V(\ve_{ij})^{(g)},
\]
where $W_A$ and $W_V$ denote the learned mappings for the attention and value branches, respectively. The attention weight is normalized over the neighboring sender cells of receiver $i$:
\[
\alpha_{ij}^{(g)}
=
\frac{
\exp\!\left(a_{ij}^{(g)}\right)
}{
\displaystyle
\sum_{j'\in\mathcal{N}_i}
\exp\!\left(a_{ij'}^{(g)}\right)
}.
\]

The influence assigned to sender $j$ for gene $g$ in receiver $i$ is then defined as
\[
I_{ij}^{(g)}
=
\alpha_{ij}^{(g)}v_{ij}^{(g)}.
\]

The level-one prediction of the receiver-cell state is obtained by directly summing these sender-specific terms:
\begin{equation}
\hat{z}_i^{(1)(g)}
=
\sum_{j\in\mathcal{N}_i}
I_{ij}^{(g)} .
\label{eq:proofsum}
\end{equation}

Equation~\ref{eq:proofsum} is the mathematical basis of the influence-tensor interpretation. Thus, the influence decomposition is exact with respect to the model prediction. It is not obtained through a post hoc attribution procedure; rather, each $I_{ij}^{(g)}$ is itself an explicit term in the forward computation of the predicted cell-state expression.

If
\[
I_i
\in
\R^{|\mathcal{N}_i|\times G}
\]
denotes the influence matrix for receiver cell $i$, then the predicted cell-state vector can be written compactly as
\[
\hat{z}_i^{(1)}
=
\mathbf{1}^{\top} I_i,
\]
where the summation is performed over the sender dimension. Each row of $I_i$ is associated with one neighboring sender cell, while each column corresponds to one target gene. Across all receiver cells, the full influence tensor
\[
I
\in
\R^{N\times k\times G}
\]
therefore retains explicit receiver, sender, and gene indices.

The attention normalization does introduce dependence on the other cells in the neighborhood through its denominator. Accordingly, $I_{ij}^{(g)}$ should be understood as the contribution assigned to sender $j$ in the context of receiver $i$'s observed neighborhood. This neighborhood dependence does not affect the exact decomposition: once the attention weights and values are computed, the individual sender-indexed terms still sum exactly to the prediction in Equation~\ref{eq:proofsum}.

\subsection*{C. Why nonlinear processing before aggregation does not invalidate the decomposition}

The pairwise encoder $F_\theta$ does not need to be linear for the influence tensor to remain interpretable. It may contain multiple transformer blocks and nonlinear transformations. These operations occur before the sender-specific influence $I_{ij}^{(g)}$ is defined and therefore determine the value of each contribution without altering the additive structure of the final prediction.

The relevant computational structure is
\[
\ve_{ij}
\longrightarrow
I_{ij}^{(g)}
\longrightarrow
\sum_j I_{ij}^{(g)}.
\]

Once the individual influence terms have been formed, the mapping from these terms to the prediction is simply summation. There is no nonlinear transformation after sender aggregation.

This distinction is important. Consider instead an architecture in which sender-specific messages $m_{ij}$ are first aggregated and the aggregated representation is subsequently transformed by a nonlinear function:
\begin{equation}
h_i
=
\sigma\!\left(
\sum_{j\in\mathcal{N}_i}
m_{ij}
\right).
\label{eq:aggthennonlin}
\end{equation}
For a general nonlinear function $\sigma$,
\begin{equation}
\sigma\!\left(
\sum_j m_{ij}
\right)
\neq
\sum_j
\sigma(m_{ij}).
\label{eq:nonadditive}
\end{equation}

The original messages $m_{ij}$ therefore cannot, in general, be interpreted as additive contributions to the final output.

A simple example illustrates this problem. Suppose two neighboring cells contribute scalar messages $m_1$ and $m_2$, and let
\[
\sigma(x)=x^2.
\]
Then
\[
\sigma(m_1+m_2)
=
(m_1+m_2)^2
=
m_1^2+m_2^2+2m_1m_2.
\]
The term
\[
2m_1m_2
\]
depends jointly on the two senders. There is no unique decomposition of this interaction term into a contribution attributable exclusively to sender 1 or sender 2. More generally, nonlinear transformations following neighborhood aggregation introduce interactions among sender messages and prevent the final prediction from being represented uniquely by the original sender-specific terms.

GITIII-scale avoids this ambiguity in its interpretable head. Nonlinear transformations are allowed while constructing each pair-specific contribution, but after the influence terms $I_{ij}^{(g)}$ are formed, the prediction is obtained directly through
\[
\hat{z}_i^{(1)(g)}
=
\sum_j I_{ij}^{(g)}.
\]
The additive relationship between the influence tensor and the prediction is therefore preserved exactly.

\subsection*{D. Why multiple message-passing layers obscure sender identity}

A second source of ambiguity in conventional GNNs arises from repeated message passing. In a multi-layer GNN, the representation of a neighboring cell at a later layer is generally no longer determined by that cell alone; it already contains information aggregated from its own neighbors.

Consider a standard message-passing architecture. At the first layer, the representation of cell $j$ may be written schematically as
\[
h_j^{(1)}
=
\sigma_1
\left(
\sum_{\ell\in\mathcal{N}_j}
m_{j\ell}^{(1)}
\right).
\]
At the second layer, receiver $i$ aggregates messages constructed from these updated representations:
\begin{equation}
h_i^{(2)}
=
\sigma_2
\left(
\sum_{j\in\mathcal{N}_i}
m_{ij}^{(2)}
\left(
h_j^{(1)},h_i^{(1)}
\right)
\right).
\label{eq:twolayer}
\end{equation}

The message associated with the edge $j\rightarrow i$ at the second layer is therefore a function of $h_j^{(1)}$. However,
\[
h_j^{(1)}
=
h_j^{(1)}
\left(
\{x_\ell:\ell\in\mathcal{N}_j\}
\right),
\]
so it already contains information from the neighboring cells of $j$. Consequently, information can reach receiver $i$ through multi-hop paths such as
\[
\ell
\longrightarrow
j
\longrightarrow
i.
\]

The apparent contribution associated with neighboring cell $j$ therefore includes information that originally came from other cells $\ell$. It can no longer be interpreted solely as the direct contribution of $j$ to $i$.

Repeated message passing can also propagate information from the same source through multiple paths. For example, a cell $\ell$ may affect receiver $i$ through both
\[
\ell
\longrightarrow
j_1
\longrightarrow
i
\]
and
\[
\ell
\longrightarrow
j_2
\longrightarrow
i.
\]
After several layers, the representation of the receiver therefore combines information transmitted along many graph paths, and the identities of the original source cells become entangled with those of the intermediate nodes.

This issue can be seen even in a simplified linear GNN. Ignoring nonlinearities, an $L$-layer graph convolution can be written schematically as
\[
H^{(L)}
=
A^L XW,
\]
where $A$ denotes a graph propagation operator. The coefficient
\[
(A^L)_{i\ell}
\]
contains contributions from the collection of length-$L$ graph walks connecting cell $\ell$ to cell $i$. Thus, the representation of receiver $i$ is inherently a multi-hop quantity rather than a direct collection of contributions from its immediate neighboring sender cells. When nonlinear transformations are inserted between the message-passing layers, these path-dependent signals are further mixed, making an exact sender-wise decomposition even less direct.

The difficulty is therefore not that multi-layer GNNs cannot learn biologically meaningful cell--cell interaction patterns. Rather, their architecture does not generally provide a direct decomposition of the final prediction into terms that remain uniquely associated with individual neighboring sender cells. Such a decomposition would require an additional attribution procedure and would no longer correspond directly to explicit terms in the forward prediction.

\subsection*{E. Interpretation}

The interpretability of the GITIII-scale influence tensor therefore follows from two complementary architectural properties. First, each sender--receiver interaction is represented separately before neighborhood aggregation, so the identity of the direct neighboring sender is retained when its influence is constructed. Second, once the sender-specific, gene-specific influences have been formed, they are aggregated only by summation, with no nonlinear transformation applied afterward.

Consequently, for every receiver cell $i$ and target gene $g$,
\[
\text{predicted cell-state expression}
=
\sum_{\text{neighboring sender cells}}
\text{sender-specific gene-level influence} ,
\]
or, equivalently, Equation~\ref{eq:proofsum}.

This equality holds exactly by construction and provides the mathematical basis for interpreting the influence tensor as an additive decomposition of the model-predicted receiver-cell state.

\end{document}